# *A Novel Framework for Intelligent Information Retrieval in Wireless Sensor Networks*

*Savneet Kaur, Deepali Virmani, Satbir Jain*

*GTBIT, BPIT, NSIT*

*GGSIPU, NSIT, Delhi, India*

*deepalivirmani@gmail.com*

*Abstract*—**Recent advances in the development of the low-cost, power-efficient embedded devices, coupled with the rising need for support of new information processing paradigms such as smart spaces and military surveillance systems, have led to active research in large-scale, highly distributed sensor networks of small, wireless, low-power, unattended sensors and actuators. While applications keep diversifying, one common property they share is the need for an efficient network architecture tailored towards information retrieval in sensor networks. Previous solutions designed for traditional networks serve as good references; however, due to the vast differences between previous paradigms and needs of sensor networks, a framework is required to gather and impart only the required information .To achieve this goal in this paper we have proposed a framework for intelligent information retrieval and dissemination to desired destination node. The proposed frame work combines three major concern areas in WSNs i.e. data aggregation, information retrieval and data dissemination in a single scenario. In the proposed framework data aggregation is responsible for combining information from all nodes and removing the redundant data. Information retrieval filters the processed data to obtain final information termed as intelligent data to be disseminated to the required destination node.**

I. INTRODUCTION TO WIRELESS SENSOR NETWORKS

A wireless sensor network commonly known as WSN is a large network, consisting of various sensors used to monitor situations like temperature, pressure, motion, pollutants, vibration etc [7]. The first use of the WSN was done in the military applications such as battlefield surveillance and now extending in almost all the day to day activities including environment, healthcare applications, home automation, traffic control, habitat monitoring and much more. The devices used in the sensor networks are small and inexpensive thus can be used and produced in large quantities. The devices' resources in terms of memory, energy, bandwidth and speed are severely controlled [11]. Wireless Sensor networks combine its three main areas namely sensing, at a single tiny device. Each device has a small microcontroller, a radio Tran receiver, and a battery for energy source. Each device gets information from other device and passes it to the monitoring computer. Wireless sensor networks have the ability to deploy large numbers of tiny nodes called the sensor nodes.

II. EXISTING FRAMEWORK

A. Existing working scenario of wireless sensor networks

Fig. 1 represents the existing architecture of sensor networks. Sensor networks consist of thousands of sensor nodes [1]. Each sensor node has the capability to collect data and send it back to the sink nodes or to the end users. A satellite link/internet is used to transfer the data from sink node to end user.

The place where the sensor nodes have to be positioned is pre-determined. These nodes function in cooperative effort. Each node has an on board processor which is used to carry out simple calculations and computational task and transmitting only the required and partially processed data.

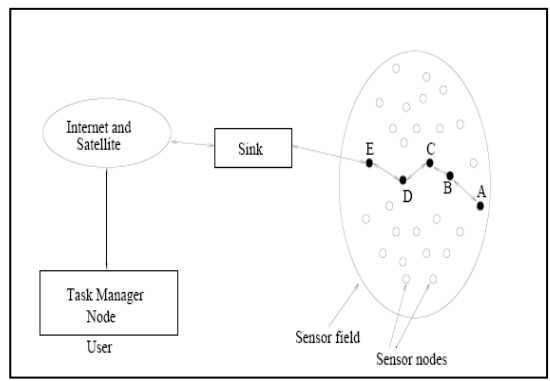

Figure 1 Sensor Network Architecture

### B. Need for new framework

Wireless Sensor Networks (WSNs) are greatly used in the applications where large and real time computation is needed. Energy consumption and communication cost are the most significant aspects of WSNs contrary to the computational cost which is less significant. Phenomena's like data aggregation and in-network processing of the data are used extensively in the field of sensor networks to increases the lifetime of such networks. Data aggregation is used to collect the data from all the nodes at a dedicated node and forward all the data through a single communication channel. Since the amount of data generated in the sensor network is usually huge therefore, we need a technology for combining this data into information at the sensor nodes so that the number of packets which has to be sent to the sink nodes be reduced. Data aggregation has the ability to reduce energy consumption and enhance the network lifetime. To fulfill the monitoring task, large scale Wireless Sensor Networks usually collect or generate vast amounts of data during their lifetime.

To segregate the data from the large data reservoirs Information Retrieval should be added with data aggregation.

Information Retrieval (IR) [10] is about the process of providing answers to users information needs. It is thus concerned with the collection, representation, storage, organization, accessing, manipulation and display, of the information items necessary to satisfying those needs.

There is need of a combined approach combining the three major aspects namely data gathering (Data Aggregation), data processing (Information retrieval) and data forwarding (Data Dissemination) [2] to the desired sink node. Existing framework/architecture only forwards all the information to the sink node which leads to energy, power and cost wastage.

So in this paper we propose a new novel framework for intelligent information retrieval in wireless sensor networks. Our proposed framework is an integrated solution to major problematic areas in WSNs i.e. minimizing energy consumption, minimizing communication cost, maximizing throughput, minimizing delay in turn maximizing network lifetime.

### III. PROPOSED FRAMEWORK

Proposed framework is presented in figure 2.For the purpose of relevant data collection and dissemination in this paper we propose a novel framework for intelligent information retrieval and dissemination. The proposed framework allows to collect the data of all the sensor nodes at a single node (sub sink), then only required information is retrieved by process of information retrieval. At the end only the final information is passed / forwarded to the sink node. So if a single channel will only forward the required/desired information then lot of energy would be conserved thus minimizing energy consumption and enhancing network lifetime.

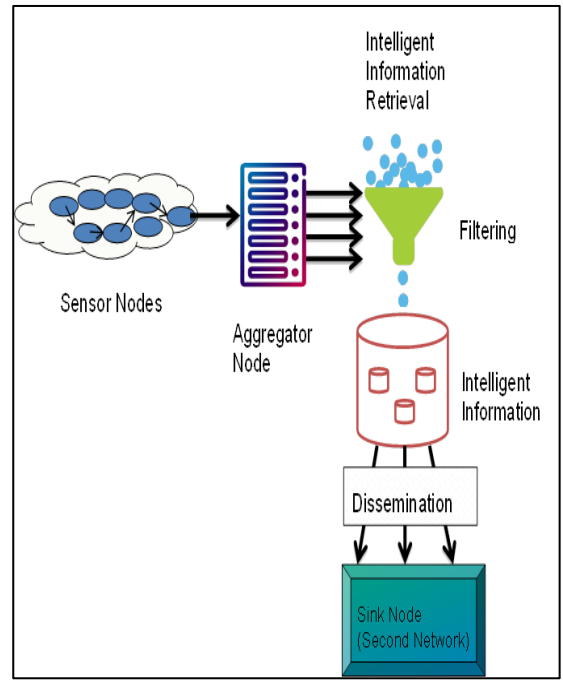

Figure 2. Proposed Framework

### A. Different stages of the proposed framework

All the data produced by the sensor nodes is forwarded to the central node called as sub-sink [8].

Sensor Nodes

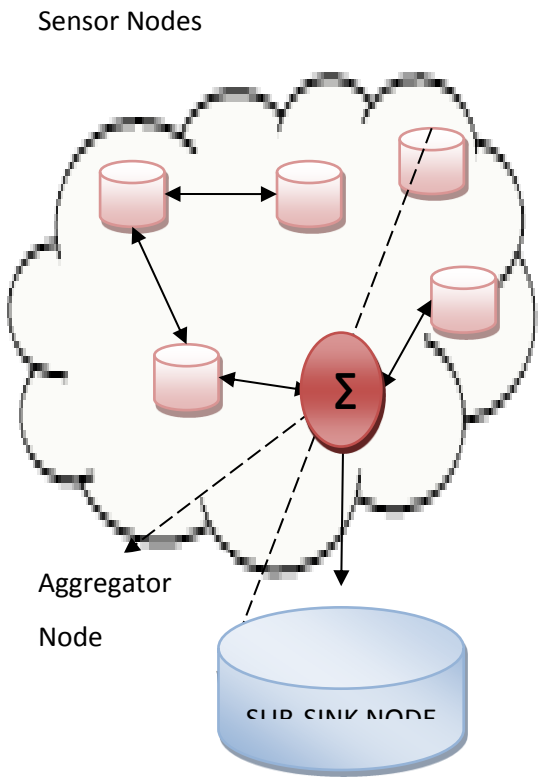

Aggregator Node

SUB SINK NODE

Figure 3 Aggregator node

i. RAW DATA: Digital components are getting cheaper day by day with the rapid growth in computer & digital technology. A sensor node is made up of these devices mainly consists of hardware, processor, memory, power supply & transmitter. Sensor network consist of lots of sensors randomly distributed in the network which are collecting & broadcasting the data. The raw data generated by sensor nodes can be processed to generate some relevant data.

ii. AGGREGATOR OF NETWORK: There are large numbers of sensor nodes existing in the sensor network which are capable of generating lots of data (figure 3). Due to this network might get chocked [11]. To overcome this problem there are some sensor nodes which can act as aggregator nodes & aggregate the data based on some computation which further can be filtered or disseminated.

iii. INTELLIGENT INFORMATION RETRIVAL: The data generated by the sensor node is collected and due to the large volume of the generated data we apply algorithms and other useful techniques to filter this data according to the given needs. Filtering process filters the data into more relevant information. Filtering is useful to increase the accuracy & decrease the redundancy in the data. Filtering operation can be achieved by indexing algorithms.

Such data after filtering is passed on to the next level of the hierarchy and is termed as intelligent information. This information is passed further to the sink nodes from where it is disseminated to the desired locations.

B. *Our proposed framework consists of the following stages*

i. Data aggregation
ii. Intelligent information retrieval
iii. Data dissemination

i. Data aggregation- In this the data is collected from all the sensor nodes and the data is worked upon to remove the redundancy [9]. Such data is forwarded for the next stage where the intelligent information from it is taken out.
ii. Intelligent Information Retrieved – The generated data is passed through a series of analysis as shown in Fig. 4 to get the intelligent information.

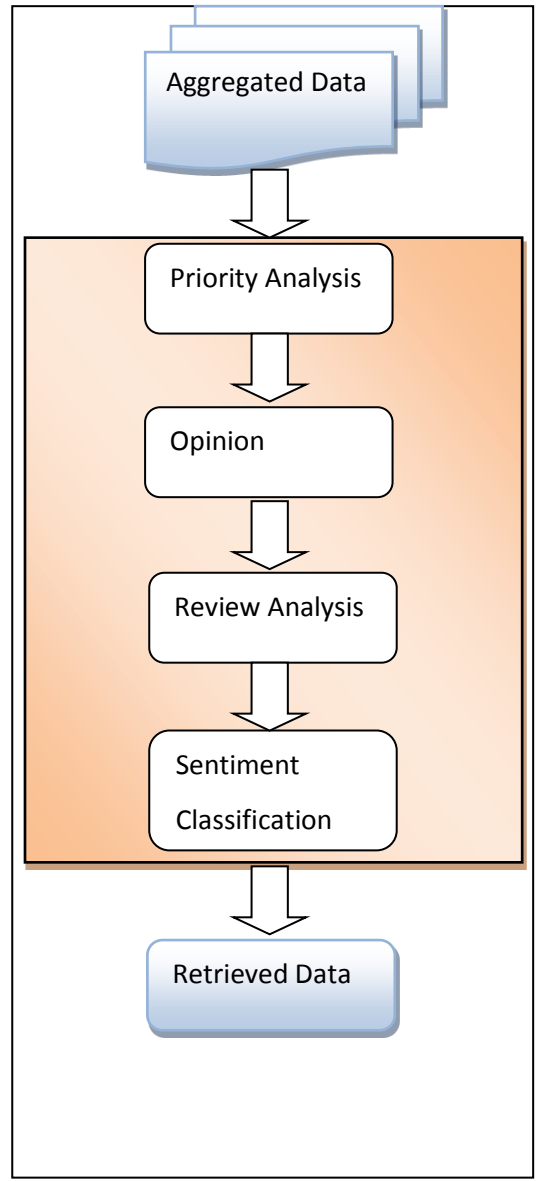

Figure. 4 Intelligent Information Retrieval Steps

To the best of our knowledge all the previous work either uses Priority Analysis or Opinion Analysis [3] or Review Analysis or Sentiment Classification [4] to retrieve the data. But in our proposed framework the information is retrieved after a proper processing by a clubbed approach. That is our proposed framework uses a staircase model, where first step of priority analysis is applied to aggregated data that is only the data with high priority is assigned task number 1 to be forwarded for dissemination.

After settling priority to data the information will be processed for further filtering by opinion analysis. By applying opinion mining to high priority data only required information is extracted based on facts and computation as only required information is the output of the Opinion Analysis phase. This information is further reviewed by applying linguistic features under the category of review analysis. Final information is filtered by symbolic analysis and supervised learning. This step is known as sentiment classification.

iii. Data Dissemination : Required filtered data is forwarded to the next node. Since the data is filtered and only the useful and the intelligent data is forwarded it leads to maximize the network lifetime by minimizing energy consumption in the network.

### C. *How is our proposed framework different from the existing?*

In our proposed framework the information retrieval is done on the basis of intelligence and the intelligence is approved by processing of data by clubbing the opinion mining, priority analysis, sentimental analysis. To the best of our knowledge the previous works have not clubbed all these techniques to retrieve the information.

## IV. ADVANTAGES OF THE PROPOSED FRAMEWORK

Information Retrieval is nothing but extracting the relevant information from the collection of document or data as per need. There is need to design good Information Retrieval system which is capable to catering user's need i.e. answers' the users query and give the relevant result. Characteristics of a good IR system are

- Faster data extraction: IR makes faster data retrieval since user does not have to go through all the document, user approach to analyze data can be based on some condition which can results in getting relevant data.

- Cost effective: A user can access the data as per his desire & process that quickly & effectively.

- Getting latest information: Because data retrieval is based on user's query, user would always get latest & updated data.

- Getting accurate information: To get the accurate information, user has to put the right query to IR system.

## CONCLUSION

In this paper we have proposed a novel framework for intelligent information retrieval in wireless sensor networks. In our proposed work we have tried to combine different types of mining approach (review analysis, opinion analysis, and sentimental review) to achieve the final information. Final processed information is final intelligent information that would be only forwarded to the sink node for dissemination. By the proposed approach we would be able to minimize the information to be forwarded that will in return minimize energy consumption leading to maximizing the network lifetime.